\documentclass[11pt,twoside]{article}
\usepackage{asp2010}

\resetcounters
\aspvolume{455}
\aspcpryear{2012}
\aspvoltitle{4$^{th}$ {\em Hinode} Science Meeting: Unsolved 
Problems and Recent Insights}
\aspvolauthor{L.~R.~Bellot Rubio, F.~Reale, and M.~Carlsson, eds.}

\begin{document}

\title{Detection of Flux Emergence, Splitting, Merging, and 
Cancellation in the Quiet Sun}
\author{Y.~Iida,$^1$ H.~Hagenaar,$^2$ and T.~Yokoyama$^1$}
\affil{$^1$University of Tokyo, Japan}
\affil{$^2$Lockheed Martin Space and Astrophysical Laboratory, USA}

\begin{abstract} We investigate the frequency of magnetic activities, 
namely flux emergence, splitting, merging, and cancellation, through
an automatic detection in order to understand the generation of the
power-law distribution of magnetic flux reported by \cite{2009ApJ...698...75P}.
Quiet Sun magnetograms observed in the \ion{Na}{i}~5896~\AA\/ line by
the {\em Hinode} Solar Optical Telescope is used in this study.  The
longitudinal fluxes of the investigated patches range from $\approx
10^{17}$~Mx to $\approx 10^{19}$~Mx.  Emergence and cancellation are
much less frequent than merging and splitting.  The time scale for
splitting is found to be $\approx 33$ minutes and independent of the
flux contained in the splitting patch.  Moreover magnetic patches
split into any flux contents with an equal probability.  It is shown
that such a fragmentation process leads to a distribution with a
power-law index $-2$.  Merging has a very weak dependence on flux
content, with a power-law index of only $-0.33$.  These results
suggest that (1) magnetic patches are fragmented by splitting,
merging, and tiny cancellation; and (2) flux is removed from the
photosphere through tiny cancellations after these fragmentations.
\end{abstract}

\vspace*{-1em}

\section{Introduction}
The surface magnetic activities are the source of various phenomena
and one of the most important targets for solar studies.
\cite{2009ApJ...698...75P} found that the magnetic flux content of the solar surface 
has a power-law distribution spanning the range from large active
regions to small patches in the quiet network that are concentrated
near supergranular boundaries.  This suggests that they are generated
by the same mechanism, or that they are dominated by the same magnetic
processes: emergence, splitting, merging, and cancellation.  The
relationship between the flux distribution and these activities is
described by the magneto-chemistry (MC) equation
\citep{1997ApJ...487..424S}. Although some special cases of the MC equation
reproduce the power-law spectra of magnetic flux content
\citep{2002MNRAS.335..389P}, there remain open questions.  Observational studies of
the frequencies of the four magnetic activities mentioned above give
important clues to this problem. Here we investigate magnetograms to
quantify the flux dependence of these magnetic activities by means of
an automatic detection. We concentrate on the quiet Sun because the
activities there are more moderate than in active regions.
 
\section{Observation}
 We use magnetograms in \ion{Na}{i} D 5896 \AA\/ obtained by the
 Narrow-band Filter Imager (NFI) of the Solar Optical Telescope (SOT)
 onboard the {\em Hinode} satellite.  The observation period is 00:30~UT --
 04:09~UT on November 11, 2009. The time cadence is 1 minute.  The
 field of view is $112\arcsec \times 112\arcsec$.  {\em Hinode} observed
 near the disk center during this period.  The upper panel of Fig.~1
 shows an example of the magnetograms.  It is possible that not all
 internetwork patches are observed even at the {\em Hinode} resolution
 \citep[see the flattened flux distribution in][]{2009ApJ...698...75P}.  We focus on
 network patches to avoid this observational limit.  Dark current and
 flat field of CCD camera are removed by using the procedure {\tt
 fg\_prep.pro} included in the {\em SolarSoft} package.  The average
 of each line is subtracted to remove a tip side effect of the CCD
 camera \citep{2010ApJ...720.1405L}.  We rotate all the magnetograms to the position
 at 02:03~UT by taking account of the solar differential rotation.
 The images were aligned to remove residual spacecraft jitter. The
 Stokes V/I signal is converted to magnetic field strength after these
 processes.  The conversion coefficient is taken to be 8645~G/DN from
 a comparison between a Stokes V/I image and an MDI magnetogram which
 were acquired within one minute.  We smear the magnetic field data
 with a scale of three pixels in space and average it over five images
 in time.

\begin{figure}
\plotone[height=19.5cm]{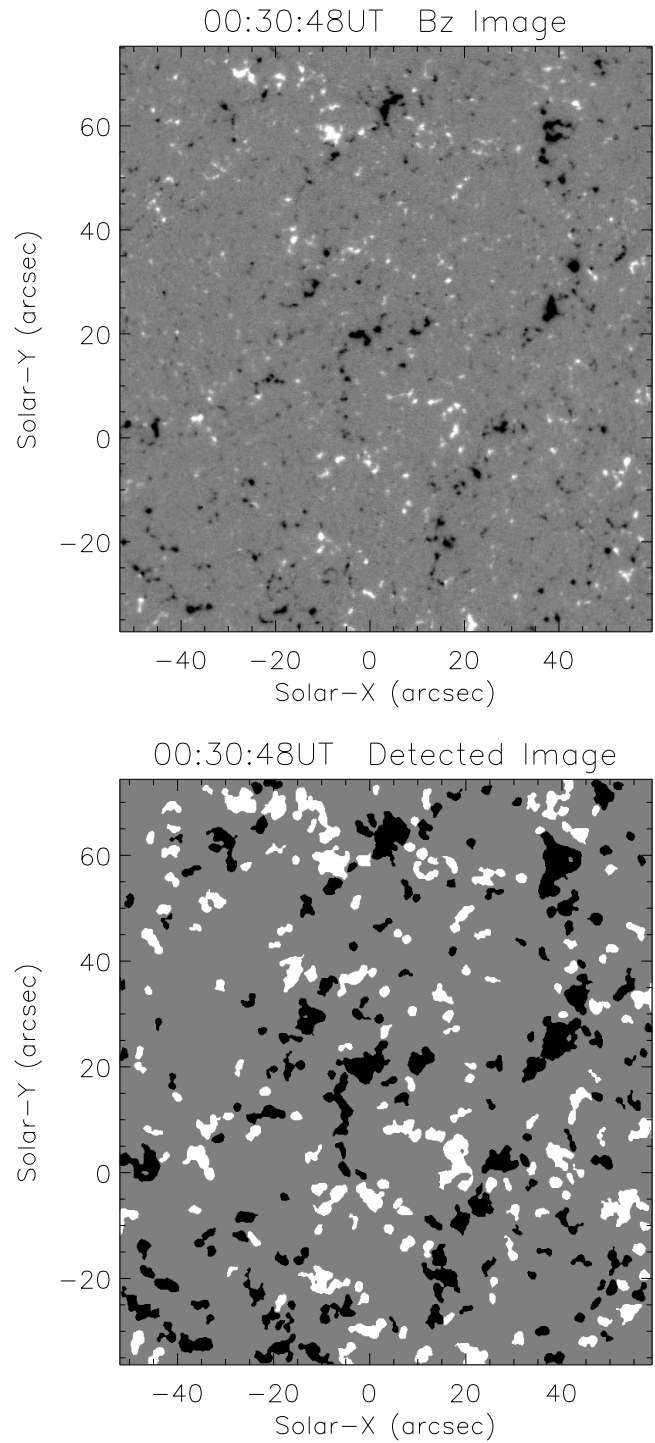}
\caption{{\em Top:} \ion{Na}{i}~D magnetogram obtained by {\em Hinode}/SOT at 
00:30~UT on November 11, 2009. {\em Bottom:} Two-valued magnetogram for
detected patches corresponding to the upper panel.}
\end{figure}

\section{Detection of Magnetic Activities}

Our procedure to investigate magnetic activities consists of (1)
detection and tracking of magnetic patches, (2) detection of
splittings and mergings, and (3) detection of emergences and
cancellations.

We detect magnetic patches by means of a clumping method, in which a
patch is defined as a massif of pixels having signals above a
threshold \citep{2009ApJ...698...75P}.  The threshold for magnetic field is set as
$2\sigma$. The value of $2\sigma$ is obtained from a Gaussian fit to
the magnetic field strength distribution and turns out to be about 
10~G.  We also set a size threshold of 81 pixels for the massifs,
which corresponds to the typical size of granules. The lower panel of
Fig.~1 shows a two-valued magnetogram computed with these thresholds.
We can see that the detected features are mainly network patches and
that most internetwork patches are removed.  Tracking is done by
checking the spatial overlap of patches in consecutive images
\citep{hag99}.  However, there are many overlaps of multiple patches
at the {\em Hinode} resolution.  Thus we set two additional conditions for
tracking: (A) tracking is done from the patches with larger flux
content, and (B) if there are more than two overlapping patches, the
one having the closest flux content is selected. These conditions
are based on the concept that smaller flux patches tend to go below
the detection threshold (condition A) and that the flux change between
consecutive images is moderate (condition B).

Detection of splittings and mergings is also done by checking the
overlap between consecutive images.  More specifically, the conditions
for splitting are the following: (C) more than two patches overlap a
patch from the previous image; and (D) more than one overlapping patch
is newly produced, i.e., it did not exist in the previous image.
Conditions C and D guarantee that one or more new patches overlap the
previous one.  Merging is defined as the same process in time-reversed
magnetograms.

Emergences/cancellations are detected as a pair of flux
increase/decrease events in different polarities after removing those
due to splittings and mergings. We define a flux change event as (E) a
flux increase or decrease for more than 5 minutes with ${\rm
d}\phi/{\rm d}t > 3.5\times10^{18}$~Mx. The threshold for $d\phi/dt$
is taken to be the typical flux change rate of cancellations because a
cancellation is thought to be a more moderate event in flux change
than an emergence \citep{2001ApJ...560..476C}.

\section{Results}

\begin{table}[t]
\caption{Number of magnetic patches and magnetic activities}
\label{tab1}
\tabcolsep 1.5em
\begin{center}
\begin{tabular}{lcc}
\noalign{\smallskip}
\tableline
\noalign{\smallskip}
&Positive&Negative\\
\noalign{\smallskip}
\tableline
\noalign{\smallskip}
Number of patches&1636&1637\\
Splitting&493&482\\
Merging&536&535\\
\noalign{\smallskip}
\tableline
\noalign{\smallskip}
Emergence&\multicolumn{2}{c}{3}\\
Cancellation&\multicolumn{2}{c}{86}\\
\noalign{\smallskip}
\tableline
\end{tabular}
\end{center}

\vspace*{-1em}
\end{table}

Table 1 shows the total number of detected magnetic patches and magnetic
activities.  There are enough splittings and mergings for a
statistical investigation but not enough emergences and cancellations.
Because splitting and merging of more than three patches occur in less 
than 5\% of the cases, they will be ignored in what follows.

Figure 2a shows the flux distribution of the detected magnetic
patches.  The dotted/broken lines indicate the distribution of
positive/negative patches, respectively, and the solid line indicates
the distribution of all the magnetic patches.  The dashed line
represents a linear fit to the flux distribution of all patches
between $10^{17.5}$~Mx and $10^{19}$~Mx. The fit indicates
$\phi^{-1.78}$, where $\phi$ is the flux content of each patch and the
error of the fitted index is $\approx 0.05$. Within the accuracy of
our results, this dependence is almost the same as the one obtained by
\cite{2009ApJ...698...75P}, $\phi^{-1.85}$. The distribution drops down 
below $\phi_{\rm{th}} = 10^{17.5}$~Mx, which we define as a detection
limit in this study. We cannot detect all the patches below the limit,
so from now on we only investigate patches with fluxes larger than
$\phi_{\rm{th}}$.

The solid line in Fig.~2b shows the flux distribution of splitting
patches divided by patch density, which means the average splitting
probability density for each patch.  We can see that the distribution
is almost flat much above the detection limit, implying that the
probability density of splitting is independent of flux content.  The
probability density of observed splittings $K(\phi)$ can be expressed
as
\begin{equation}
K(\phi)=\int_{0}^{\phi}k(x,\phi-x) \ {\rm d}x=\rm{constant} \; \; (\rm{for\ all}\ \phi),
\end{equation}
where $k(x,y)$ is the splitting probability density distribution into
flux content $x$ and $y$. Around the detection limit, the probability
density is falling down.  This falling down is explained by splitting
into fluxes below the detection limit.  We need another condition of
splitting besides Equation 1---how the flux content is split---to
estimate the dropping around the detection limit.  We assume that
patches split into elements of any flux content with equal
probability, namely
\begin{equation}
k(x,\phi-x)=\rm{constant} \ (\rm{for}\ \it{x}).
\end{equation}

\begin{figure}[p]
\plotone[height=15.7cm]{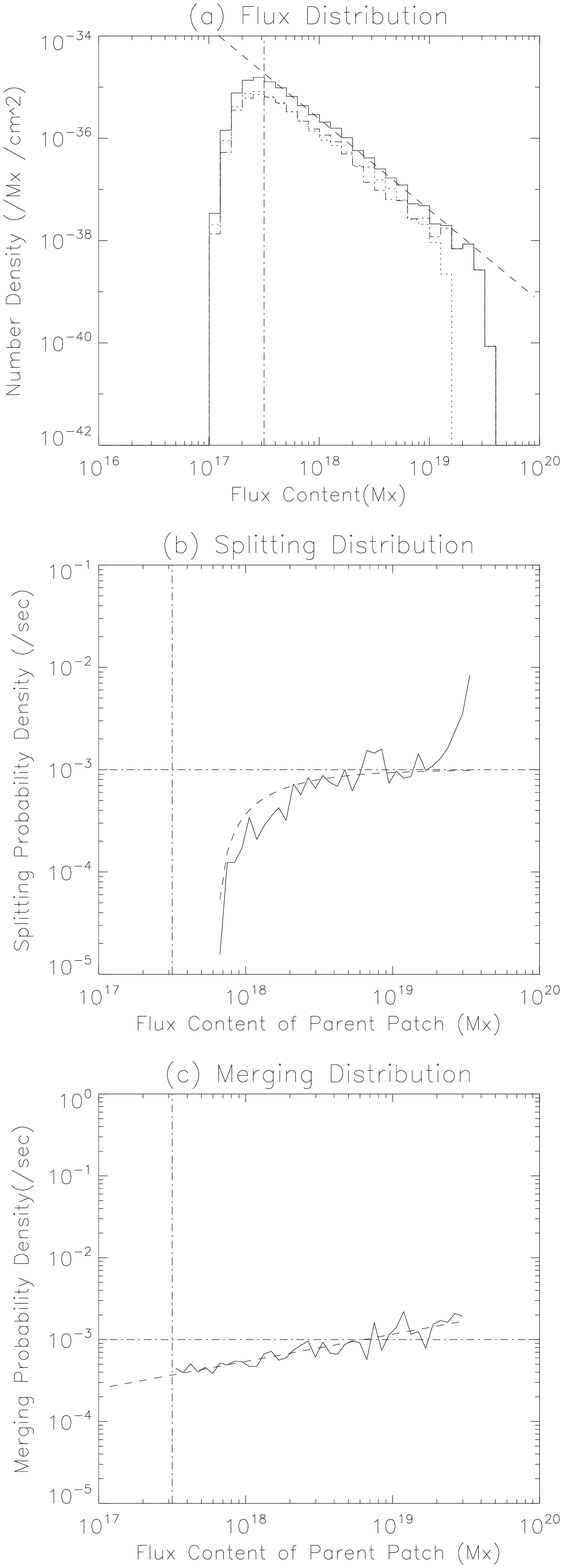}
\caption{{\em (a):} Distribution of magnetic flux contained in magnetic patches.
The solid/dotted/broken lines indicate the observed flux distribution
of all/positive/negative patches, respectively. The dashed line
represents a linear fit between $10^{17.5}$~Mx and $10^{19}$~Mx. The
slope of the fitted line is $-1.78$, showing nearly the same power law
distribution reported by \cite{2009ApJ...698...75P}. The vertical line indicates the
lower limit for the validity of the power law distribution in this
study, which corresponds to the detection limit ($=10^{17.5}$~Mx).
{\em (b):} Observed probability density distribution of splitting flux
patches (solid) and an analytical probability density distribution for
random splitting with a timescale of $33$ minutes (dashed).  The
horizontal and vertical lines indicate the probability densities
corresponding to $33$ minutes and the detection limit.  {\em (c):} Observed
probability density distribution of merging flux patches (solid) and
linear fit (dashed).  The slope of the fitted line is $0.33$. The
vertical line marks the detection limit.  }
\end{figure}

Equations 1 and 2 lead to the following analytic form for the probability
density distribution of flux content for splitting patches
\begin{equation}
k(x,y)=\frac{K_0}{x+y},
\end{equation}
where $K_{0}$ represents the typical time constant for splittings.
Now we can estimate the falling down effect of splitting into flux below
the detection limit as
\begin{equation}
K(\phi; \phi_{\rm{th}}) = \int_{\phi_{\rm{th}}}^{\phi-\phi_{\rm{th}}}k(x,\phi-x)\, {\rm d}x = K_0 \, (1-\frac{\phi_{\rm{th}}}{\phi}).
\end{equation}
The dashed curve in Fig.~2b indicates $K(\phi; \phi_{\rm{th}})$ with
$K_0=1.0\times10^{-3}$~$\rm{s}^{-1}$.  This estimated line fits the
falling down of the observed distribution well, which supports the
assumption that patches split into any flux content with equal
probability.

Figure 2c shows the flux distribution of merging flux. The solid
line represents the observed distribution normalized by patch density,
which indicates the probability of merging of one patch.  There are no
effect of $\phi_{\rm{th}}$ because the flux content after merging is
larger than before the merging.  The dashed line indicates a fit of
the form $\phi^{-0.33}$.

\section{Discussion}

\subsection{Power-law Distribution Generated by Splitting Process}

By using the obtained dependence of splitting on flux content, we
derive the flux distribution achieved through the accumulation of
this process.  The MC equation for splitting can be written as
\begin{equation}
\frac{\partial n(\phi)}{\partial t} = 2\int_{\phi}^{\infty}n(x)k(\phi,x-\phi)\, 
{\rm d}x - \int_{0}^{\phi}n(\phi)k(x,\phi-x)\, {\rm d}x
\end{equation}
where $n(\phi)$ represents the flux distribution.  After substituting
$k(x,y)=K_0/(x+y)$ and differentiating with respect to $\phi$, we
obtain
\begin{equation}
\frac{\partial^2n(\phi)}{\partial \phi \partial t} = - \frac{K_0}{\phi^2}\frac{\partial}{\partial \phi}\left[\phi^2n(\phi)\right].
\end{equation}
This equation has a time-independent solution $n(\phi) \propto \phi^{-2}$.

\subsection{Importance of Tiny Cancellations}

We note here that cancellation may play an important role in the
generation of the observed power-law flux distribution. Figure~3 shows
a schematic view of our model. For the flux distribution we
consider a power law of index $-\gamma$ which, according to the
observations, is in the range $1.5<\gamma<2$.  The total flux amount
can be calculated as
\begin{equation}
\Phi_{\rm tot} = \int_{\phi_{\rm min}}^{\phi_{\rm max}} \phi \, n(\phi) \, {\rm d}\phi \propto \phi_{\rm max }^{-\gamma+2},
\end{equation}
where the rightmost side is valid for $\gamma < 2$.  This shows that
the total flux is dominated by $\phi_{\rm max}$.  We obtain the total
flux loss by cancellations as
\begin{equation}
\left. \frac{\partial \Phi_{\rm tot }}{\partial t} \right|_{\rm canc }=\int_{\phi_{\rm{min}}}^{\phi_{\rm{max}}}\phi \frac{\partial n_{\rm{canc}}(\phi)}{\partial t}\, {\rm d}\phi
\propto \phi_{\rm{min}}^{-2\gamma+3}.
\end{equation}
This implies that tiny cancellations play an important role in flux balance.

\begin{figure}[p]
\epsscale{0.99}
\plotone[height=7cm]{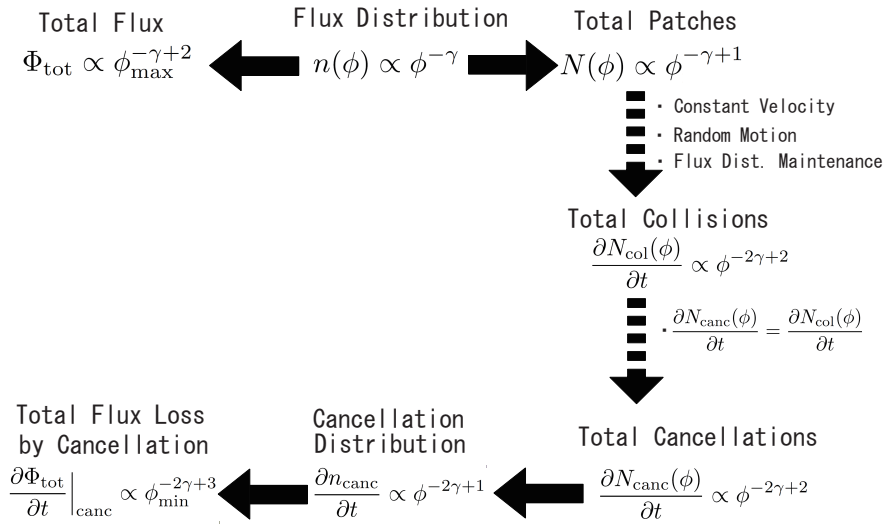}
\caption{A model for the flux distribution of magnetic patches 
and cancellations with $1.5<\gamma<2$.  Solid arrows indicate
mathematical relations, while dashed arrows indicate relationships
with some physical assumptions.  Note that the total flux is dominated
by $\phi_{\rm{max}}$ but that the total flux loss is dominated by
$\phi_{\rm{min}}$ in this model.}
\end{figure}

\begin{figure}[p]
\plotone[height=9.5cm,bb=0 70 572 640]{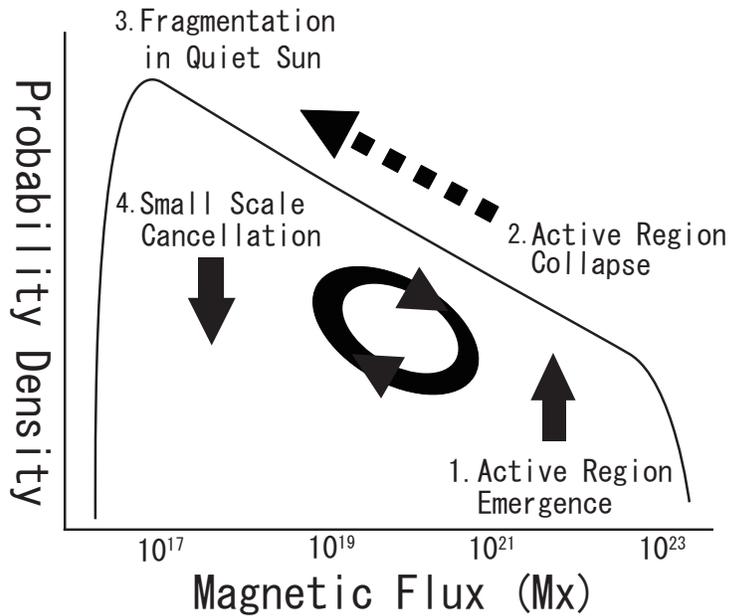}
\caption{A schematic picture of the flux transport mechanism. 
The total flux is dominated by large emerged fluxes. Emerged fluxes
are fragmented through splittings, mergings, and cancellations with
tiny magnetic patches.  Cancellations become dominant in tiny patches
(see Fig.~3 and related discussion).}
\end{figure}

\subsection{Global Flux Transport Speculation}
In Figure~4 we speculate on the flux transport on a global scale. The
magnetic flux from active regions is mainly fragmented by pure
splittings and cancellations with tiny patches.  They will submerge
through cancellations when they are fragmented enough.  These
speculations should be confirmed in the future with a full
understanding of the four magnetic activities discussed in the present
paper.

\acknowledgements
We thank all the members of the {\em Hinode} team and the GCOE program
`From the Earth to Earths', which supported a stay in Lockheed Martin
Space and Astronomical Laboratory. We also thank the members of
Lockheed Martin Space and Astronomical Laboratory for plentiful
discussions.

\bibliography{hinode4}

\end{document}